\documentclass[12pt,fleqn]{article}
\usepackage{verbatim,amssymb,amsfonts,amsmath,graphics,amsthm,cite}
\numberwithin{equation}{section}

\newtheorem{theorem}{Theorem}[section]

\def\p{{\vec{p}}}
\def\s{{\mathsf{s}}}

\def\u{{\vec{u}}}

\def\H{{\cal H}}

\def\P{{\cal P}}

\def\<{\langle}
\def\>{\rangle}

\def\l{\left|\!\left|}

\def\r{\right|\!\right|}
\begin{document}
\title{On Einstein Causality and Time Asymmetry in Quantum Physics}
\author{S.~Wickramasekara and A.~Bohm\\
Department of Physics, University of Texas at Austin\\
Austin, Texas 78712\\
E-mail: sujeewa@physics.utexas.edu}
\date{}
\maketitle
\begin{abstract}
A theorem of Hegerfeldt shows that if the spectrum of the Hamiltonian
is bounded from below, then the propagation speed of certain
probabilities does  not have an upper bound.
We prove a theorem analogous to 
Hegerfeldt's that appertains to 
asymmetric time evolutions given by a semigroup of operators. As an
application, we consider a characterization of 
relativistic quasistable states by irreducible representations of the
causal Poincar\'e semigroup and study the implications of the new
theorem for this special case. 
\end{abstract}

\section{Introduction}\label{sec1}
G.~C.~Hegerfeldt discovered some interesting features of the structure
of quantum physics appertaining to microphysical causality
\cite{h1,h4}.  In particular, he showed that an initially
localized particle immediately develops ``infinite tails'', unless it
remains localized for all times. This is a  
very broad and general result in that it does not depend on the
details specific to a particular quantum system such as the form of
the Lagrangian and type of interaction. Only the existence of a
positive selfadjoint 
Hamiltonian, and therewith the symmetry of time translations, is assumed.

In its final form, the discovery of Hegerfeldt is encapsulated in the
following theorem:
\begin{theorem}\cite{h4}\label{th1}
Let $H$ be a selfadjoint operator, positive or bounded from below, in
a Hilbert space $\H$. For given $\psi_0\in\H$, let $\psi_t,\
t\in{\mathbb{R}}$, be defined as $\psi_t=e^{-iHt}\psi_0$. Let A be a
positive operator\footnote{It seems to us that Hegerfeldt's proof
holds only when $A$ is a {\em bounded\/} positive operator, 
though this is not
explicitly stated in \cite{h4}.} in $\H$ and $p_A(t)$ be defined as
$p_A(t)=\<\psi_t,A\psi_t\>$. Then, either $p_A(t)\not=0$ for almost
all $t$ and the set of such $t$'s is open and dense, or  $p_A(t)=0$
for all $t$.
\end{theorem}

It is a classical result due to Stone that every selfadjoint operator
$H$ leads to a strongly continuous one parameter group of unitary operators
$U(t)=e^{-iHt}$. If the spectrum of the operator $H$ is bounded from
below, then the mapping $t\rightarrow U(t)$ admits an analytic
extension into the open 
lower half of the complex plane, ${\mathbb{C}}^-$, i.e., the
mapping $z\rightarrow U(z)=e^{-iHz}$ is strongly analytic for every
$z$ with $\Im{z}<0$. Further, since $z_1+z_2\rightarrow
U(z_1+z_2)=U(z_1)U(z_2)$ and $\l U(z)\r\leq1$, the analytic mapping
$z\rightarrow U(z)$ furnishes a representation of the semigroup
${\mathbb{C}}^-$ (under addition) by contractions in $\H$. The proof
of Theorem \ref{th1} is anchored in this analytic extension of
$e^{-iHt}$ into the semigroup $e^{-iHz},\ z\in{\mathbb{C}}^-$.

In the next section, we shall show that a result analogous to Hegerfeldt's
theorem holds for certain quantum systems the time evolution of which
is given by a semigroup of bounded normal operators. Specifically, we
shall prove  
that $p_A(t)=\<\psi_t,A\psi_t\>\not=0$ for almost all $t\geq0$, unless
$p_A(t)\equiv0$ for all $t\geq0$. The condition $t\geq0$, rather than
$t\in{\mathbb{R}}$ as in Hegerfeldt's result, is a consequence of the
fact that the time evolution is now furnished by a semigroup --not a
unitary group as required for Theorem \ref{th1}. 

Semigroup time evolutions are generally understood as representing a
time asymmetry or an irreversibility at the quantum physical level. 
The main theoretical question given rise to by our mathematical result
is whether such asymmetry in time translations implies or is consistent
with Einstein causality. The immediate emergence of ``infinite tails''
for an initially localized particle, a prediction of Hegerfeldt's
theorem, remains a feature of the particular semigroup time evolution
developed in the next section. Thus, insofar as particle localization 
is concerned, {\em if} such localization for a relativistic particle
is meaningful in the first place, 
it appears that asymmetry in time evolution and causality are quite
distinct notions --in particular, the former does not ensure the latter.   

However, to further examine the issues of Einstein causality for
irreversible processes, we must also consider the space translations
consistent with the semigroup time evolution (i.e., the set of
spacetime translations resulting from all possible boosts of the time
evolution semigroup). The characterization of relativistic quasistable
states by irreducible representations of a particular subsemigroup of
the Poincar\'e group, studied in Section 3, provides a concrete
example with such a semigroup of spacetime translations. Within the
context of this example, we discuss aspects of Einstein causality
inferred by the main mathematical result of this paper.

\section{A Variant of Hegerfeldt's Theorem}
The following variant of Theorem \ref{th1} holds:

\begin{theorem}\label{th2}
Let $\H$ be a Hilbert space, and $H$, a normal operator in $\H$. 
Suppose the following conditions hold on $\sigma(H)$, the spectrum of
$H$:
\begin{eqnarray}
\sup_{\lambda\in\sigma(H)} (\lambda_y)&=&k_0<\infty\label{2.1}\\
\sup_{\lambda\in\sigma(H),\ \lambda_x\leq0}
\left(\frac{-\lambda_y}{\lambda_x}\right)&=&k_1<0\label{2.2}\\
\inf_{\lambda\in\sigma(H),\ \lambda_x\geq0}
\left(\frac{-\lambda_y}{\lambda_x}\right)&=&k_2>0\label{2.3}
\end{eqnarray}
where $\lambda=\lambda_x+i\lambda_y$.\\
Let $e^{-iHt},\ t\in[0,\infty)$, be the semigroup generated by $-iH$,
and let $p_A(t)$ be defined as
\begin{equation}
p_A(t)=\<\psi_t,A\psi_t\>\label{2.4}
\end{equation}
where $A$ is a bounded selfadjoint operator in $\H$, and
$\psi_t=e^{-iHt}\psi_0$ for given $\psi_0\in\H$. Then,\\ 
either
\begin{equation}
p_A(t)\not=0 \ \text{for almost all}\ t\in[0,\infty)\label{2.5}
\end{equation}
and such $t$'s are open and dense in $(0,\infty)$\\
or
\begin{equation}
p_A(t)=0\ \text{for all}\ t\in[0,\infty)\label{2.6}
\end{equation}
\end{theorem}

\noindent{\bf Proof.}
It is well known \cite{rudin} that if $\sigma(H)$ satisfies
\eqref{2.1}, the operator $(-iH)$ generates a strongly continuous one
parameter 
semigroup of bounded normal operators $e^{-iHt},\
t\in[0,\infty)$. Conditions \eqref{2.2} and \eqref{2.3} ensure that
the strongly continuous mapping $t\rightarrow e^{-iHt}$ admits an
extension into the domain $D=\{z=t+iy:\ t>0,\ k_1<y<k_2\}$ such that
it is strongly continuous on $D$ and strongly analytic on
$D\setminus(0,\infty)$.  

The dual semigroup $e^{iH^\dagger t}$ admits an extension
into $D'=\{z=t+iy:\  t>0,\ -k_2<y<-k_1\}$. As before, the extension is
continuous on $D'$ and analytic on $D'\setminus(0,\infty)$ in the strong
operator topology. Therefore, the function
$p_A(t)=\<\psi_t, A\psi_t\>=\<\psi_0, e^{iH^\dagger
t}Ae^{-iHt}\psi_0\>$ has an extension into
$D\cap D'=\{z=t+iy:\ t>0,\ -k<y<k\}$, where $k=\min\{|k_1|,
k_2\}$.

This extension, which we denote by $p_A$, 
is continuous on $D\cap D'$ and analytic on
$(D\cap D')\setminus(0,\infty)$. Further, since $\overline{p_A(\bar{z})}=p_A(z)$
for every $z\in(D\cap D')\setminus(0,\infty)$ and $p_A(t)$ is real for
$t\in(0,\infty)$, by the Schwarz principle of reflection, $p_A$ is
analytic in $(D\cap D')$.    
Since $(0,\infty)$ is in the domain of analyticity of
$p_A$, either \eqref{2.5} or \eqref{2.6} must hold.\hfill$\Box$\\

\noindent{\bf Remark:} Notice that unlike in Theorem \ref{th1}, 
the positivity of $A$ is  not
necessary for our proof. 

\section{Application to Relativistic Decaying States}\label{sec3}

Wigner's pioneering work established that a correspondence
exists between the unitary irreducible representations of the
Poincar\'e group $\P$ and relativistic (stable) particles by way of
their mass and spin. In the Hilbert space $L_{mj}^2({\mathbb{R}}^3)$
of momentum wavefunctions for a particle of mass $m$ and spin $j$, the
relevant unitary representation can be realized by
\begin{equation}
(U(\Lambda, a)f)(\p,j_3)=e^{-ip_\mu
a^\mu}\sum_{j'_3}D^j_{j'_3j_3}(W(\Lambda,p))f
(\vec{\Lambda^{-1}p},j'_3)\label{3.1}
\end{equation}
where $(\Lambda,a)\in\P$,  
$\{\p\}={\mathbb{R}}^3$ and $p^2=m^2$. The 
$W(\Lambda,p)$ are the well known
Wigner rotations. The $D^j$ matrices
provide a $(2j+1)$-dimensional representation of the quantum
mechanical rotation subgroup.  

The unitary irreducible representation for the particle of mass $m$
and spin $j$ can be realized also in the Hilbert space
$L^2_j({\mathbb{R}}^3)$ of ``four-velocity wavefunctions'':
\begin{equation}
(U(\Lambda,a)f)(\u,j_3)=e^{-imu_\mu
a^\mu}\sum_{j'_3}D^j_{j'_3j_3}(W(\Lambda,u))f
(\vec{\Lambda^{-1}u},j'_3)\label{3.2}
\end{equation} 
where $u_\mu=\frac{p_\mu}{m}$. The crucial property that makes
\eqref{3.2} possible is that Wigner rotations $W(\Lambda,p)$ are functions
only of the quotients $\frac{p_\mu}{m}=u_\mu$, but not of the $p_\mu$
themselves~\cite{werle}. That is, $W(\Lambda,p)=W(\Lambda,u)$. The inner product under which
the velocity wavefunctions form the Hilbert space
$L^2_j({\mathbb{R}}^3)$ is
\begin{equation}
(f,g)=\sum_{j_3}\int\frac{d^3\u}{2u_0}\overline{f(\u,j_3)}g(\u,j_3)\label{3.3}
\end{equation}
where $\{\u\}={\mathbb{R}}^3$ and $j_3=-j, -j+1,\cdots,j$.

It is generally believed that relativistic resonances and unstable
particles are associated with simple poles of an analytic
$S$-matrix. The location of the pole, $\s_R=(M-i\Gamma/2)^2$, contains
information about the ``mass'' and ``width'' of the particular
quasistable state\footnote{We use the term ``quasistable state'' to refer to
resonances and unstable particles collectively.} while the specific
partial wave in which the quasistable state occurs as a pole defines its
spin $j$. Much like the unitary irreducible representations
\eqref{3.2} of $\P$ which 
characterize isolated stable particles, we may define a representation for a
quasistable state of (complex) square mass $\s_R$ and spin $j$  in the
Hilbert space $L^2_j({\mathbb{R}}^3)$ by
\begin{equation}
(U(\Lambda, a)f)(\u,j_3)=e^{-i\sqrt{\s_R}u_\mu
a^\mu}\sum_{j'_3}D^j_{j'_3j_3}(W(\Lambda,u))f
(\vec{\Lambda^{-1}u},j'_3)\label{3.4}
\end{equation}
where, as in \eqref{3.2}, $\{\u\}={\mathbb{R}}^3$ and $j_3=-j,
-j+1,\cdots,j$. 

Notice that the operators $U(\Lambda, a)$ defined by \eqref{3.4} are bounded
in $L^2_j({\mathbb{R}}^3)$ if and only if $(\Lambda, a)\in\P_+$, where
\begin{equation}
\P_+=\{(\Lambda,a):\ (\Lambda, a)\in\P,\ a_0\geq0,\ 
a^2\geq0\}\label{3.5}
\end{equation}
The subset $\P_+\subset\P$ is a subsemigroup of $\P$ as it remains
invariant under the product rule $(\Lambda_1,a_1)(\Lambda_2,a_2)=(\Lambda_1\Lambda_2,
a_1+\Lambda_1a_2)$ for $\P$. Further, the inverse mapping
$(\Lambda,a)\rightarrow(\Lambda,a)^{-1}=(\Lambda^{-1},-\Lambda^{-1}a)$ transforms
every element $(\Lambda,a)\in\P_+$ (except $(I,0)$) out of $\P_+$. The
$\P_+$ is the semidirect product of the group of 
orthochronous Lorentz transformations 
and the semigroup $T_+$ of spacetime translations into the
forward light cone:
\begin{equation}
T_+=\{a:\ a_0\geq0,\ a^2\geq0\}\label{3.6}
\end{equation}

The operators defined by \eqref{3.4} provide a continuous
representation of $\P_+$ by contractions in
$L_j^2({\mathbb{R}}^3)$. This representation 
is characterized by a real, discrete
spin value $j$ and a complex square mass value $\s_R=(M-i\Gamma/2)^2$,
where $M$ and $\Gamma$ may be interpreted as the mass and width of the
quasistable state, respectively. Notice further that, as in \eqref{3.2},
the Lorentz  
subgroup of $\P_+$ is unitarily represented by \eqref{3.4}. It must be
mentioned that complex mass representations with real
velocities of the Poincar\'e transformations 
have been considered in the past~\cite{zwanziger}.

The Hamiltonian $H=P_0$ for \eqref{3.4} 
acts in $L_j^2({\mathbb{R}}^3)$ as
\begin{equation}
(P_0f)(\u,j_3)=\sqrt{\s_R}u_0f(\u,j_3),\quad
u_0=(1+\u^2)^{1/2}
\end{equation}
This shows that $P_0$ is a normal operator whose spectrum
$\sigma(P_0)=\{(M-i\Gamma/2)u_0:\ u_0\in[1,\infty)\}$.
Thus, $\sigma(P_0)$ satisfies the conditions demanded in Theorem
\ref{th2}, and we have a semigroup of time evolution
\begin{equation}
(U(t)f)(\u,j_3)=e^{-i\sqrt{\s_R}u_0t}f(\u,j_3)
=e^{-iM(1+\u^2)^{1/2}t}e^{-\frac{\Gamma}{2}(1+\u^2)^{1/2}t}f(\u,j_3),
\quad t\geq0\label{3.9} 
\end{equation}

Now, if $f_t=e^{-iP_0t}f$ for any $f\in L_j^2({\mathbb{R}}^3)$ and $A$
is any bounded selfadjoint operator in $L_j^2({\mathbb{R}}^3)$, then
by Theorem \ref{th2}, $p_A(t)=\<f_t,Af_t\>$ is either almost never
zero for $t\geq0$ or identically zero. 
In particular, if we choose for $A$ a projection operator $N(V)$
representing the localization of the quasistable particle in a finite
volume $V$, a calculation can be carried out in complete analogy to
that in \cite{h4} to show that there appear ``infinite tails'' for any
$t\geq0$. That is, the asymmetric time evolution of the quasistable
state does not preclude the propagation of its survival probability
into all space for arbitrarily small positive times. However, unlike
Hegerfeldt's theorem, ours does not yield non-zero probabilities for
$t<0$ since the time evolution is now given by a semigroup with
$t\geq0$. The mathematical result notwithstanding, it is important
here to mention that the spatial localization of a relativistic
quasistable state, as given by the above projection operator $N(V)$,
may not be a physically meaningful notion. It has been argued
\cite{buchholz} that, even for a relativistic stable particle, the
localization by way of the projection $N(V)$ considered in \cite{h4}
does not have much physical content.

What may be relevant and more meaningful as a directly measurable
quantity (unlike survival probability) is the decay probability of the
quasistable state into its possible decay products. Thus, if the
operator $A$ taken to be the projection representing the measurement
of certain decay products, then \eqref{2.4} and \eqref{2.5} infer that
the probability to detect decay products becomes non-zero immediately
after the quasistable state is created at $t=0$. This property,
however, does not say anything about the Einstein causality for the
spacetime propagation of decay probability, a problem which requires
that we consider the whole semigroup \eqref{3.6} of spacetime
translations. To that end, recall first that
the operators $e^{-iP_\mu a^\mu}$ of \eqref{3.4}
are bounded in $L_j^2({\mathbb{R}}^3)$ only when $a\in T_+$. 
For such $a\in T_+$, 
\begin{equation}
(U(a)f)(\u,j_3)=e^{-i\sqrt{\s_R}u_\mu a^\mu}f(\u,j_3)
=e^{-i\sqrt{\s_R}(u_0t-\u.\vec{x})}f(\u,j_3)\label{3.10}
\end{equation}
where $a=(t,\vec{x})$. 

Consider now a certain fixed $\vec{x}$. By \eqref{3.10}, the 
$U(a)$ is a bounded operator only when $t\geq|\vec{x}|$. Further, the
mapping $a\rightarrow U(a)$ given by \eqref{3.10} admits an extension
into $D=\{z=t+iy:\ t>|\vec{x}|,\
y<\frac{\Gamma}{2M}(t-|\vec{x}|)\}$ such that it is strongly
continuous in $D$ and strongly  analytic in
$D\setminus(|\vec{x}|,\infty)$. Starting from this observation, the
proof technique of
Theorem \ref{th2} can be invoked to conclude that, for any bounded
operator $A$ in $L_j^2({\mathbb{R}}^3)$  and given $f\in
L_j^2({\mathbb{R}}^3)$, the quantity 
$p_A(t,\vec{x})=\<U(a)f, AU(a)f\>$ is either almost never zero for 
$t\geq|\vec{x}|$ or identically zero. 

Either of these two possibilities is clearly causal, and so what
remains to be examined is $p_A(t,\vec{x})$ in spacelike
regions. Ideally, we expect a causal theory to predict vanishing
expectation values $p_A(t,\vec{x})=0$ in all regions of spacetime
outside the forward semi-lightcone $T_+$. While the theory presented
here does not yield this ideal result, certain properties of the
transformation operators $U(a)$ for $a\not\in T_+$ motivate the
plausible argument that the theory is mathematically meaningful only
for $a\in T_+$. In particular, the operators $U(a)$ are unbounded for
$a\not\in T_+$, and for such $a$, the following results are true:
\begin{itemize}
\item[{i)}] There exist elements $f$ of $L^2_{mj}({\mathbb{R}}^3)$ such
that $||U(a)f||$ is infinite. 
\item[{ii)}] For any $f\in L^2_{mj}({\mathbb{R}}^3)$ and $\alpha>0$,
there exists some $a\not\in T_+$ such that $||U(a)f||>\alpha$. That
is, the function $a\rightarrow ||U(a)f||$ grows without bound for any
$f\in L^2_{mj}({\mathbb{R}}^3)$. 
\end{itemize}
If we want to maintain the general idea that the vectors $f$ and
$U(a)f$ represent the same physical system as viewed by two different
observers, then the transformations $U(a)$ for $a\not\in T_+$ need to
be excluded from the theory. In particular, if $U(a)$ for $a\not\in
T_+$  are admitted, then it implies that for every normalized state
$f$ one can find an observer, spacelike separated from the first, to
whom the quantity $||U(a)f||$ can be arbitrarily large. That is, there
is no natural way to attribute a probability interpretation for $U(a)f$
when $a\not\in T_+$.  Likewise, for a state $f$ that is initially
normalized, i.e., $||f||=1$, the quantity  $p_A(t,\vec{x})=\<U(a)f,
AU(a)f\>$ cannot be interpreted as the expectation value of the
observable $A$  in an arbitrary state $f\in L_j^2({\mathbb{R}}^3)$ at
$a=(t,\vec{x})$, if $a\not\in\P_+$. It may be possible\footnote{This
was pointed out by a referee.} that finite
values can be restored if the expectation values are defined as
$p_A(t,\vec{x})=\frac{\<U(a)f, AU(a)f\>}{\<U(a)f,U(a)f\>}$. This,
however, implies that the state vector is re-normalized for each
$a\not\in T_+$, and how such an operation can be given an unambiguous
physical meaning is not clear.

\section{Concluding Remarks}\label{sec4}

Theorem \ref{th1} proved by Hegerfeldt seems to imply, on the face of
it, that if nature admits only positive (or bounded from below) energies,
then its behavior is non-causal at the microphysical level. It may be
possible to resolve this 
conflict, as the author suggests, once field theoretic
concepts such as vacuum fluctuations and 
weak causality are introduced. Mathematically, the centrally
significant component of Hegerfeldt's result is the
semi-boundedness of the Hamiltonian and the resulting 
unitary group which determines the time evolution of stationary quantum
systems represented in a Hilbert space. It is this reversible group
evolution which brings about the non-vanishing probabilities for
(almost) all $t$, both positive and negative.

On the other hand, it has been argued that certain quantum mechanical
processes such as decay exhibit asymmetric,
irreversible time evolutions~\cite{lee}. Such asymmetric time evolutions can be
described by semigroups. The main technical result we reported in this
paper is a variant of Hegerfeldt's theorem that applies for 
certain asymmetric time evolutions given by semigroups.  
The characterization of unstable
particles by the complex mass representations of the semigroup $\P_+$
provides a concrete example where the new theorem can be
applied. The semigroup representation \eqref{3.4} characterized by
$\s_R$ and $j$, leads to non-zero probabilities $p_A(t)$ for (almost)
all $t\geq0$. Negative times are excluded from the prediction by
virtue of the semigroup time evolution defined only for
$t\geq0$.  

To examine the problem of Einstein causality for the
spacetime evolution of the expectation values $p_A(t,\vec{x})$ of an
observable $A$, it is necessary to consider, in
addition to the time evolution semigroup, the set of space translations
consistent with this semigroup (i.e., those obtained by boosting the
asymmetric time translations). It was seen in Section \ref{sec3} that
$p_A(t,\vec{x})$ may be  given a clean, unambiguous interpretation
only in non-spacelike regions if the
spacetime translations of the quantum system are defined by
\eqref{3.4}. Thus, while we have not obtained Einstein causality in
the stronger sense, i.e., $p_A(t,\vec{x})=0$ for all $a\not\in T_+$, in
the foregoing weaker sense, the characterization of the quasistable
states by irreducible representations of the Poincar\'e semigroup
appears to infer Einstein causality for the expectation values of decay
products.

\section*{Acknowledgment}
We acknowledge the financial support from the Welch Foundation.

\end{document}